\documentclass[twocolumn,showpacs]{revtex4}
\usepackage{graphicx}
\usepackage{bm}

\begin{document}

\title{Central and tensor components of three-nucleon forces 
\\in low-energy proton-deuteron scattering}
\author{S. Ishikawa} \email[E-mail:]{ishikawa@i.hosei.ac.jp}
\author{M. Tanifuji}
\affiliation{Department of Physics, Science Research Center, Hosei University, 
2-17-1 Fujimi, Chiyoda, Tokyo 102-8160, Japan}
\author{Y. Iseri}
\affiliation{Department of Physics, Chiba-Keizai College, 
4-3-30 Todoroki-cho, Inage, Chiba 263-0021, Japan}

\date{\today}

\begin{abstract}
Contributions of three-nucleon forces (3NF) to proton-deuteron scattering observables at energies below the deuteron breakup threshold are studied by solving the Faddeev equation that includes the Coulomb interaction. 
At $E_p=3.0$ MeV, we find that the central part of a two-pion exchange 3NF removes the discrepancy between measured cross sections and the calculated ones by two-nucleon forces, and improves the agreement with $T_{22}$ experimental data. 
However, the tensor part of the 3NF fails in reproducing data of the analyzing power $T_{21}$ by giving worse agreement between the measured and the calculated. 
Detailed examinations of scattering amplitudes suggest that a $P$-wave contribution in spin quartet tensor amplitudes has unsuitable sign for reproducing the $T_{21}$ data.
\end{abstract}

%

\maketitle


Contributions of three-nucleon forces (3NF) have been studied extensively for proton-deuteron ($pd$) scattering observables, since the interactions are successful \cite{Sa86} in solving the problem of three-nucleon (3N) underbinding for realistic two-nucleon forces (2NF). 
However, the Faddeev calculation, which is one of the practical methods to treat 3N systems, has conventionally neglected the Coulomb interaction because of mathematical difficulties, while the interaction is essentially important in the low-energy $pd$ scattering. 
Recently the problem has been solved by an Faddeev integral equation approach at energies below the deuteron breakup threshold \cite{Is03}, where the phase-shift parameters by the calculation agree to those by an Faddeev differential equation approach \cite{Fr96,Ki01} as well as the Kohn variational method \cite{Ki96,Ki01} with very good accuracy. 
This allows us to investigate the contribution of 3NF in the low-energy $pd$ scattering by the Faddeev calculation including the Coulomb interaction. 
In this Rapid Communication, we will report the main result of such calculations, where one will find remarkable effects of the 3NF with promising success of the central part but pessimistic failure of the tensor one.

In the calculation, we adopt the Argonne $V_{18}$ model (AV18) \cite{Wi95} for the 2NF and the Brazil model (BR) \cite{Co83} for the two-pion exchange 3NF. Further we introduce two kinds of 3NF: a spin-independent Gaussian (GS) 3NF \cite{Is03}
\begin{equation}
V_\textrm{GS-3NF} = V_0^G 
 \sum_{i \ne j \ne k} \exp\{-(\frac{r_{ij}}{r_G})^2 -(\frac{r_{ki}}{r_G})^2\}
\label{eq:V_GS}
\end{equation}  
with $V_0^G=-40$ MeV and $r_G=1.0$ fm; a spin-orbit (SO) 3NF \cite{Ki99}
\begin{equation}
V_\textrm{SO-3NF} =\frac12 W_0 \exp\{ -\alpha\rho\} 
  \sum_{i > j} [\bm{l}_{ij} \cdot (\bm{\sigma}_i+\bm{\sigma}_j)] 
   \hat{P}_{11},
\label{eq:V_SO}
\end{equation}
where $\rho^2=\frac23(r_{12}^2+r_{23}^2+r_{31}^2)$, $\alpha=1.5$ fm${}^{-1}$, $W_0=-20$ MeV,  and $\hat{P}_{11}$ is the projection operator to the spin and isospin triplet state of the $(i,j)$ pair.
Calculated $^3$He binding energies are 7.79 MeV for the BR-3NF in addition to the AV18 (AV18+BR), 7.74 MeV for the GS-3NF (AV18+GS), and 7.74 MeV for the BR- and SO-3NFs (AV18+BR+SO), which are compared with the empirical value of 7.72 MeV. 
The GS-3NF simulates the central part of the BR-3NF \cite{Is02}, and then the difference between both calculations describes the contribution of the tensor part of the BR-3NF. 
The SO-3NF is adopted as a simulation of the spin-vector type 3NF that reproduces empirical vector analyzing powers, but the origin is yet unknown at present. 

\begin{figure}[t]
\includegraphics[scale=0.45]{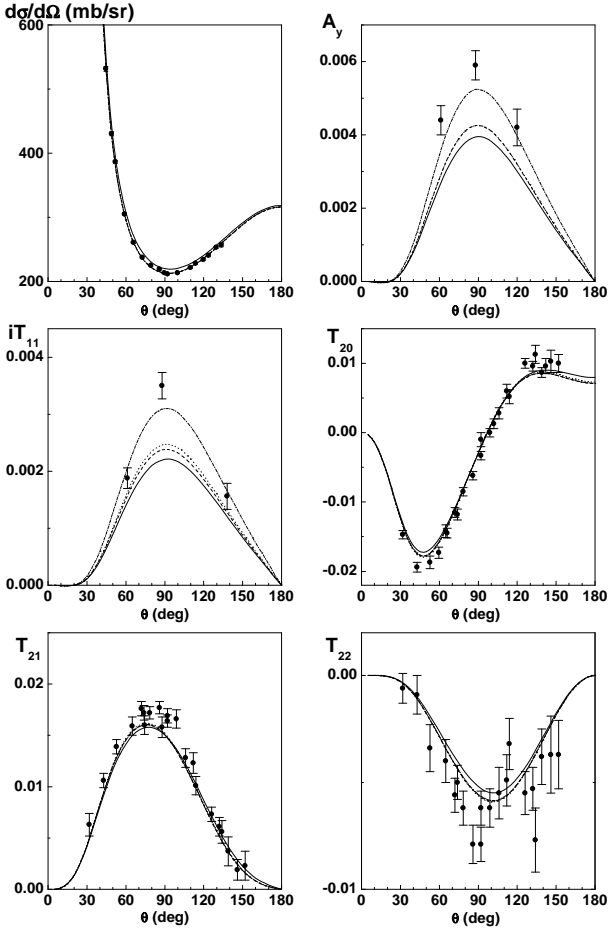}
\caption{\label{fig1}
Differential cross sections $d\sigma/d\Omega$, vector analyzing powers $A_y$ and $iT_{11}$, and tensor analyzing powers $T_{20}$, $T_{21}$, and $T_{22}$ of the $pd$ scattering at $E_p=0.65$ MeV. 
The data are taken from Ref.\ \protect\cite{Br01}.
The solid, dashed, dotted. and dash-dotted lines are the calculations by the AV18, the AV18+BR, the AV18+GS, and the AV18+BR+SO, respectively. }
\end{figure}

\begin{figure}[t]
\includegraphics[scale=0.45]{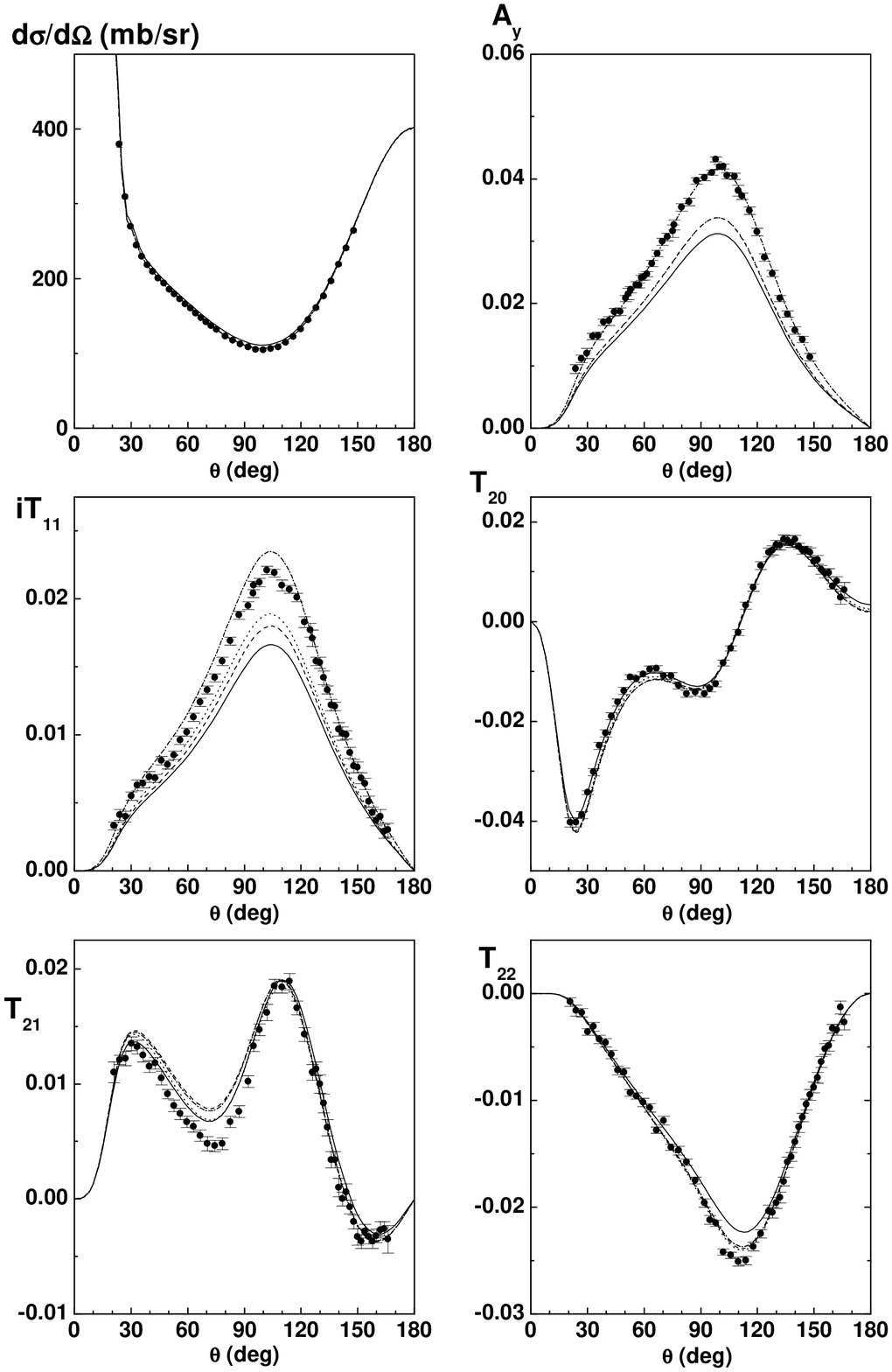}
\caption{\label{fig2}
Same as Fig.\ \protect\ref{fig1} but at $E_p=2.5$ MeV.
The data are taken from Refs.\ \protect\cite{Sa94,Sh95}.
}
\end{figure}

\begin{figure}[t]
\includegraphics[scale=0.45]{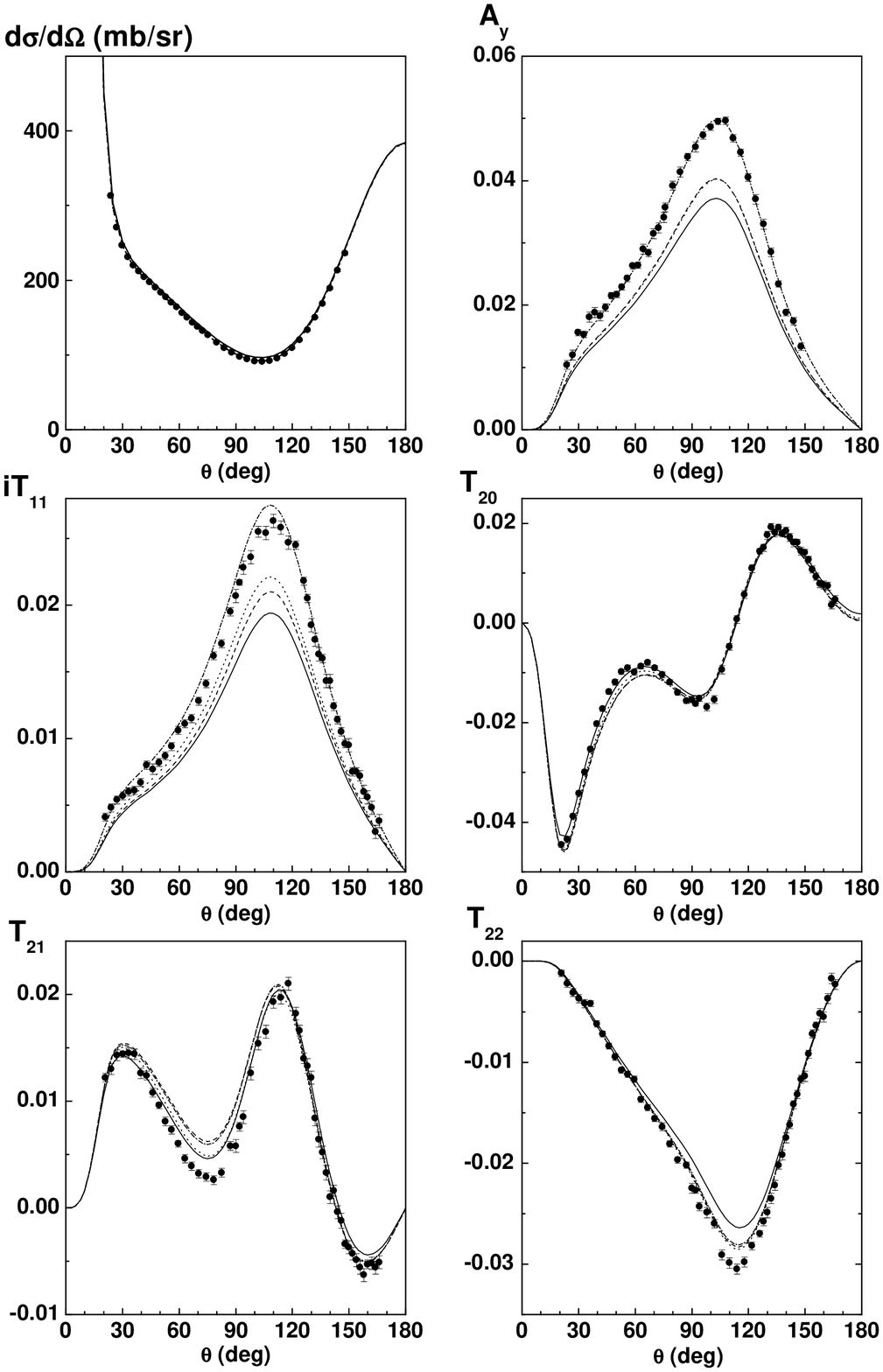}
\caption{\label{fig3}
Same as Fig.\ \protect\ref{fig1} but at $E_p=3.0$ MeV.
The data are taken from Refs.\ \protect\cite{Sa94,Sh95}. 
}
\end{figure}

The comparison between the calculated quantities and the measured ones \cite{Sa94,Sh95,Br01} is shown in Figs.\ \ref{fig1}, \ref{fig2}, and \ref{fig3} for $E_p=$ 0.65, 2.5, and 3.0 MeV, respectively, where the differential cross section $d\sigma/d\Omega$, the vector analyzing power of the proton $A_y$, that of the deuteron  $iT_{11}$, and the tensor analyzing powers $T_{20}$, $T_{21}$, and $T_{22}$ are displayed. 
Although overall agreements are obtained between the calculations and the experimental data at the three incident energies, the BR-3NF seems to deteriorate the agreement to the $T_{21}$ data of middle angles at $E_p=$ 2.5 and 3.0 MeV. 
This effect can be confirmed by $\chi^2$/datum in Table\ \ref{table1}, where we include similar analysis of the data \cite{Wo02} at $E_p=$ 1.0 MeV to see energy dependence of $\chi^2$ in detail.

\begin{table}
\begin{center}
\caption{\label{table1}
$\chi^2/datum$ of the AV18, AV18+GS, AV18+BR, and AV18+BR+SO $pd$ observables compared with the experimental data from Refs.\ \protect\cite{Br01,Sa94,Sh95,Wo02}.
}
\begin{tabular}{lcccccc}
\hline\hline
 & $\sigma$ & $A_y$ & $iT_{11}$ & $T_{20}$ & $T_{21}$ & $T_{22}$ \\
\hline
$E_p$=0.65 MeV \\
AV18       & 21  & 16 & 13 & 2.9 & 3.4 & 2.7 \\ 
AV18+GS    & 1.7 & 12 & 7.8 & 2.3 & 2.7 & 2.0 \\
AV18+BR    & 2.0 & 12 & 9.5& 2.6 & 3.1 & 2.1 \\
AV18+BR+SO & 1.7 & 2.4 & 1.5 & 2.6 & 3.2 & 2.1 \\
\hline
$E_p$=1.0 MeV \\
AV18       & 51  & 164 & 60 & 3.7 & 3.8 & 4.2  \\ 
AV18+GS    & 2.8 & 115 & 28  & 1.4 & 2.8 & 1.5 \\
AV18+BR    & 2.8 & 116 & 37 & 1.6 & 3.4 & 1.7 \\
AV18+BR+SO & 2.5 & 20  & 2.1 & 1.8 & 3.5 & 1.8 \\
\hline
$E_p$=2.5 MeV \\
AV18    & 23 & 189 &  90 & 1.5 & 5.2 & 10 \\ 
AV18+GS & 1.9 & 111 & 33 & 3.7 &  5.0 &  2.6  \\
AV18+BR & 1.8 & 111 & 53 & 6.3 & 10 & 2.8 \\
AV18+BR+SO & 2.2 & 1.9 & 10 & 5.5 & 8.9 & 3.1 \\
\hline
$E_p$=3.0 MeV\\
AV18 & 25 & 160 & 114 & 3.2 & 7.1 & 18.0 \\ 
AV18+GS & 2.3 & 95 & 47 & 5.1 & 8.3 & 3.6 \\
AV18+BR & 2.2 & 94 & 74 & 10 & 20 & 4.3 \\
AV18+BR+SO & 2.6 & 1.9 & 6.0 & 9.0 & 18 & 5.3 \\
\hline\hline
\end{tabular}
\end{center}
\end{table}

\begin{figure*}[t]
\includegraphics[scale=0.5]{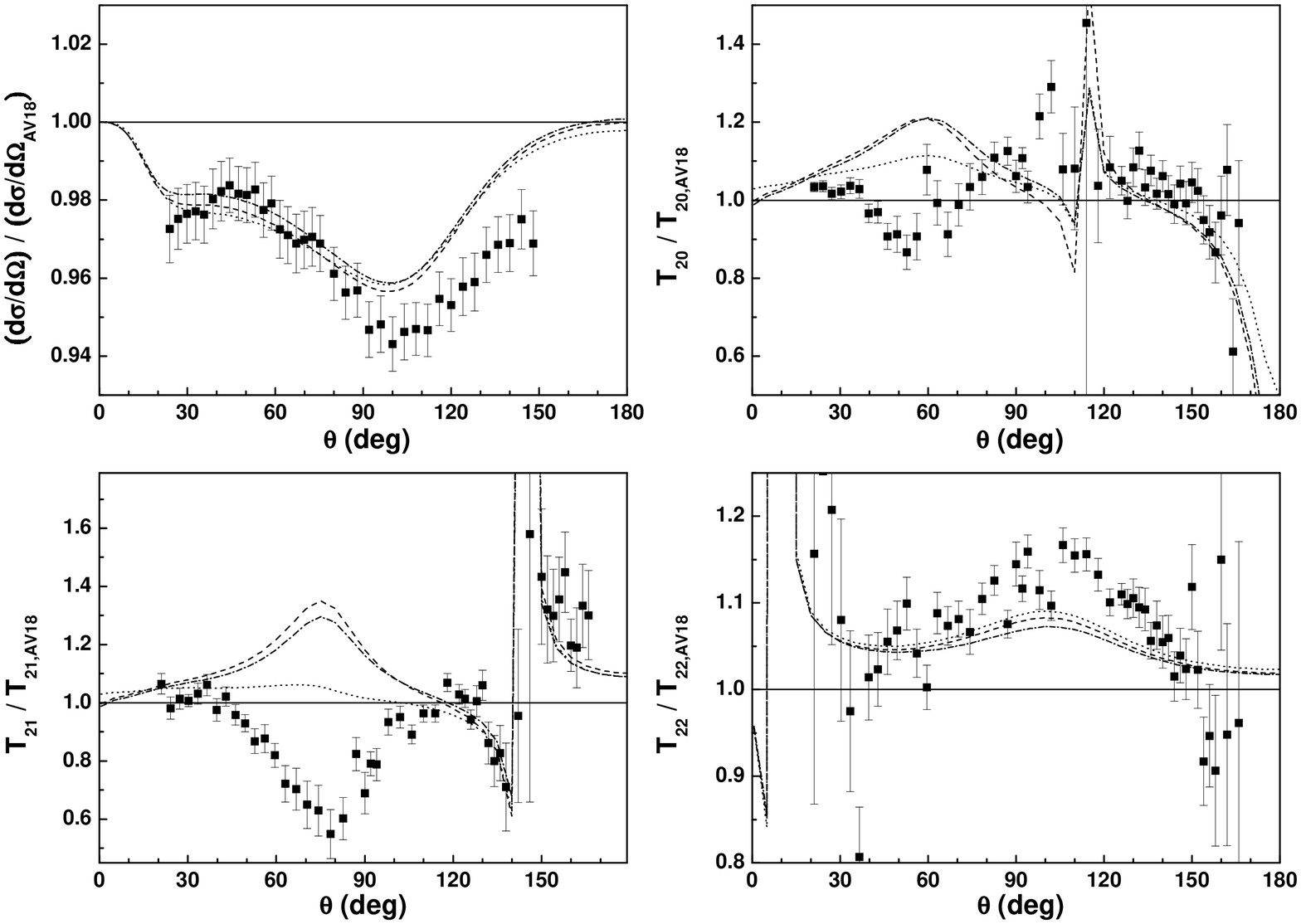}
\caption{\label{fig4}
Differential cross sections $d\sigma/d\Omega$ and tensor analyzing powers $T_{20}$, $T_{21}$, and $T_{22}$ at $E_p = 3.0$ MeV divided by the theoretical values with the AV18. 
See the caption of Fig.\ \protect\ref{fig1} for the definitions of the theoretical curves.}
\end{figure*}

To demonstrate characteristic features of such BR-3NF contributions, we will show, in Fig.\ \ref{fig4}, $d\sigma/d\Omega$, $T_{20}$, $T_{21}$, and $T_{22}$ typically at $E_p=3.0$ MeV, divided by the theoretical values obtained by the AV18 calculations. 
In the figure, the solid horizontal lines (the 2NF-lines) describe the AV18 calculations, and the deviations of the theoretical curves from the 2NF-lines describe the 3NF contributions. 
The experimental points of $d\sigma/d\Omega$ deviate from the 2NF-line with characteristic angular distribution, and the deviation is well reproduced by the calculations including the 3NF effects. 
This 3NF effect is attributed to the central part of the BR-3NF, because the calculations by the AV18+BR and those by the AV18+GS give similar results, and the additional SO-3NF produces very small contributions. 
As was discussed in Ref.\ \cite{Is02}, this 3NF contribution is produced through the spin doublet scalar amplitude and is related to the 3NF contribution in the 3N binding energy \cite{Is99}. 

Contrary to such success, the BR-3NF fails in reproducing the experimental data of $T_{21}$. 
Fig.\ \ref{fig4} shows that $T_{21}$ calculated by the AV18+BR at middle angles is located in the opposite side of the measured values with respect to the 2NF-line, indicating the opposite sign of the 3NF contribution to be desirable for reproducing the data. 
This failure is dominantly due to the tensor part of the BR-3NF, since the contributions of the GS-3NF and the SO-3NF are small. 
Similar discrepancy is also observed in $T_{20}$, but the role of the tensor part is not clear. 
The agreement with the $T_{22}$ data is improved, though is not sufficient, by taking account of the 3NF, where the dominant contribution of the 3NF comes from the central part of the BR-3NF since three calculations by the AV18+BR, the AV18+GS, and the AV18+BR+SO calculations give similar results.

Next we will examine the 3NF contributions in more detail for $T_{21}$ and $T_{22}$. 
To understand roles of the central interaction and the tensor one of the 3NF individually, we will decompose the scattering amplitude according to the tensor property in the spin-space by expanding the T-matrix $\bm{M}$ into the spin-space tensors $\bm{S}^{(K)}_{\kappa}$ \cite{Is02}
\begin{equation}
\bm{M}=\sum_{K \kappa}
  (-)^{\kappa} \bm{S}^{(K)}_{-\kappa} \bm{R}^{(K)}_{\kappa},
\label{eqs1}
\end{equation}
where $\bm{R}^{(K)}_{\kappa}$ is the coordinate-space tensor and $K$ $(\kappa)$ is the rank ($z$ component) of the tensor. 
Then the scattering amplitude is given by Ref.\ \cite{Ta68} as
\begin{eqnarray}
&&\langle \nu_p^{\prime} \nu_d^{\prime};{\bm{k}}_f | \bm{M} 
  | \nu_p \nu_d;{\bm{k}}_i \rangle 
\nonumber \\
 &=& \sum_{s_i s_f} (s_ps_d\nu_p\nu_d|s_i\nu_i) 
  (s_ps_d\nu_p^{\prime}\nu_d^{\prime}|s_f\nu_f) (-)^{s_f-\nu_f}
\nonumber\\
& & \times \sum_K (s_i s_f \nu_i -\nu_f | K \kappa) 
  M^{(K)}_{\kappa} (s_i s_f)
\label{eq3}
\end{eqnarray}      
and
\begin{eqnarray}
  M^{(K)}_{\kappa}(s_i s_f) &=&
 \sum_{\ell_i=\bar{K}-K}^K
  \left[ C_{\ell_i}(\hat{k}_i) \otimes C_{\ell_f=\bar{K}-\ell_i}(\hat{k}_f)\right]^{(K)}_{\kappa}
\nonumber \\
 && \times F^{(K)}(s_i s_f \ell_i),
\label{eq:M_K_kappa}
\end{eqnarray}
where $\bar{K} = K$~$(K+1)$ for $K=$ even (odd), 
and $C_{\ell m}(\hat{k})$ is related to $Y_{\ell m}(\hat{k})$ as usual \cite{Is02}. 
The quantity $s_i$ ($s_f$) denotes the channel spin in the initial (final) state, which is $\frac12$ (the spin doublet states) or $\frac32$ (the spin quartet states). 
The function $F^{(K)}(s_i s_f \ell_i)$, the invariant amplitude, is a function of the scattering angle $\theta$ and is designated by the tensor rank $K$.
Thus the amplitude  describes the scattering by interactions classified by $K$, i.e., $F^{(0)}(s_i s_f \ell_i)$ ($F^{(2)}(s_i s_f \ell_i)$) describes the scattering by the central (tensor) interactions. 
Because of the time reversal theorem 
\begin{equation}
F^{(K)}(s_f s_i l_f) = (-)^{s_i-s_f} F^{(K)}(s_i s_f l_i),
\label{eq7}
\end{equation}
only five amplitudes of nine tensor ones are independent \cite{Is02}.

\begin{figure}[t]
\includegraphics[scale=0.45]{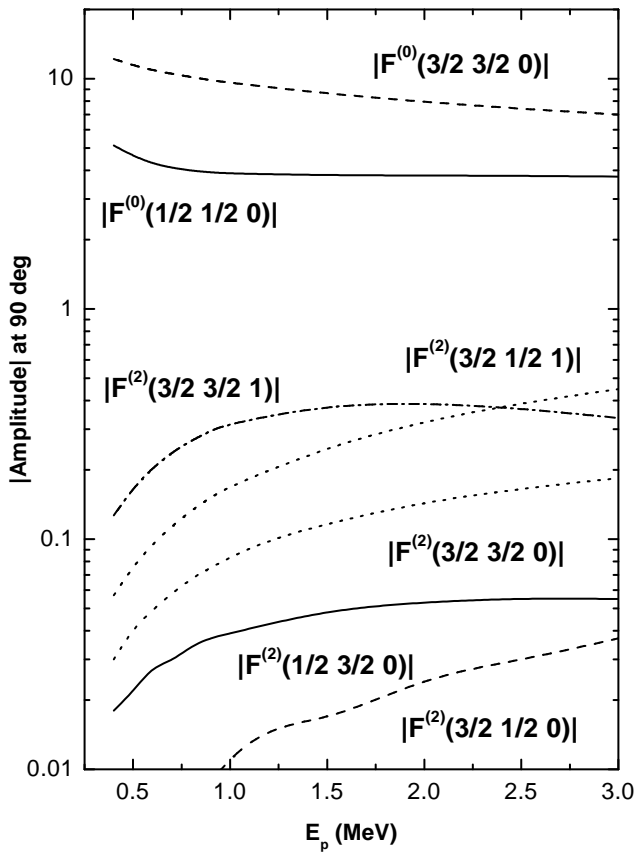}
\caption{\label{fig5}
Magnitudes of the invariant amplitudes $F^{(0)}(\frac32~\frac32~0)$, 
$F^{(0)}(\frac12~\frac12~0)$, 
$F^{(2)}(\frac32~\frac32~0)$, $F^{(2)}(\frac32~\frac12~0)$, 
$F^{(2)}(\frac12~\frac32~0)$, 
$F^{(2)}(\frac32~\frac32~1)$, and $F^{(2)}(\frac32~\frac12~1)$ 
at $\theta=90^\circ$ as functions of $E_p$ calculated with the AV18.}
\end{figure}

In the low-energy scattering, the scalar amplitudes dominate other amplitudes as shown in Fig.\ \ref{fig5}, where the magnitudes of the two scalar and five independent tensor amplitudes at $\theta=90^{\circ}$ are displayed as functions of $E_p$. 
Thus we will apply to the tensor analyzing powers $T_{2\kappa} (\kappa=0, 1,2)$ an approximation in which terms not including the scalar amplitudes are neglected, and get the following expressions:
\begin{equation}
T_{2\kappa}=T_{2\kappa}^{[1]} +T_{2\kappa}^{[2]}+T_{2\kappa}^{[3]},
\label{eq:T2kappa}
\end{equation}
where
\begin{eqnarray}
T_{2\kappa}^{[1]}&=&\frac1{N_R} \textrm{Re} \left\{-2M^{(0)}_0(\frac12 \frac12)^{*}M^{(2)}_{\kappa}(\frac32\frac12)\right\},
\label{eq:T2k_1}
\\
T_{2\kappa}^{[2]}&=&\frac1{N_R} \textrm{Re} \left\{\sqrt2 M^{(0)}_0(\frac32 \frac32)^{*}M^{(2)}_{\kappa}(\frac12 \frac32)\right\},
\label{eq:T2k_2}
\\
T_{2\kappa}^{[3]}&=&\frac1{N_R} \textrm{Re} \left\{\sqrt2 M^{(0)}_0(\frac32 \frac32)^{*}M^{(2)}_{\kappa}(\frac32 \frac32)\right\},
\label{eq:T2k_3}
\end{eqnarray}
with
\begin{equation}
N_R={\textrm Tr}(\bm{M} \bm{M}^{\dagger}) = 6\frac{d\sigma}{d\Omega}.
\label{eq12}
\end{equation}

\begin{figure}[hbt]
\includegraphics[scale=0.3]{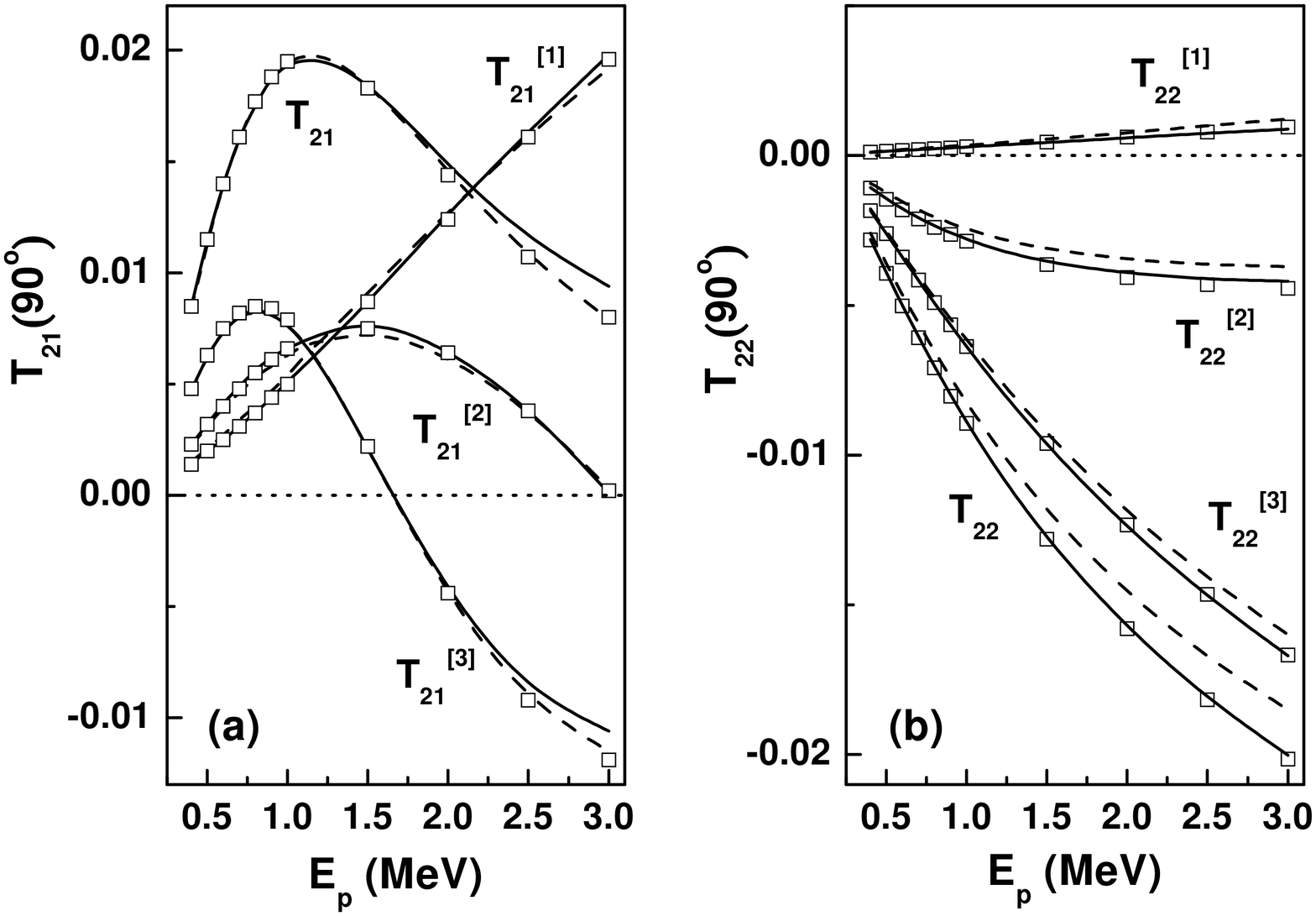}
\caption{\label{fig6}
Contributions of $T_{2\kappa}^{(1)}(90^\circ)$, $T_{2\kappa}^{(2)}(90^\circ)$, and $T_{2\kappa}^{(3)}(90^\circ)$ to $T_{2\kappa}(90^\circ)$ as functions of $E_p$ for $\kappa=1$ (a) and $\kappa=2$ (b). 
The dashed lines, the solid ones, and the square blocks represent the calculations by the AV18, the AV18+BR, and the AV18+GS, respectively.}
\end{figure}

Now our attention will be focused on the special scattering angle $\theta=90^{\circ}$ for simple considerations. 
Each component and the sum in Eq.\ (\ref{eq:T2kappa}) at $\theta=90^{\circ}$ are  displayed for $T_{21}$ ($T_{22}$) in Fig.\ \ref{fig6}(a) (Fig.\ \ref{fig6}(b)) as functions of $E_p$.
Fig.\ \ref{fig6}(a) shows that the unfavorable effect of the BR-3NF on $T_{21}$ seen in Fig.\ \ref{fig4} appears for $E_p > 2$ MeV, and  the largest contribution arises from $T_{21}^{[3]}$ of the spin quartet scattering. 
Since the quartet scalar amplitude $M^{(0)}_0(\frac32 \frac32)$ is hardly affected by the 3NFs as shown in Ref.\ \cite{Is02}, the spin quartet tensor amplitude $M^{(2)}_1(\frac32 \frac32)$ included in $T_{21}^{[3]}$ should be responsible for the effect.  
Adopting the Madison convention  for the coordinate system, $\hat{z} \parallel \bm{k}_i$ and $\hat{y} \parallel \bm{k}_i \times \bm{k}_f$, one can write down $[C_{\ell_i}(\hat{k}_i) \otimes C_{\ell_f}(\hat{k}_f)]^{(2)}_{1}$ as functions of $\cos\theta$ and $\sin\theta$. 
Referring to Eq.\ (\ref{eq:M_K_kappa}), we can restrict the invariant amplitude effective for $M^{(2)}_1 (s_i s_f)$ to $\ell_i=\ell_f=1$ ($P$-wave) as long as we are concerned with $\theta=90^{\circ}$, since the amplitudes of $(\ell_i, \ell_f) = (0, 2)$ vanish because of the factor $\cos\theta\sin\theta$, and those of $(\ell_i, \ell_f) = (2, 0)$ do not appear. 
In Eq.\ (\ref{eq:M_K_kappa}), $\ell_i$ $(\ell_f)$ gives the orbital angular momentum in the initial (final) channel, when the $\theta$-dependence of $F^{(K)}(s_i s_f \ell_i)$ is neglected because of low-energy scattering. 
Then we conclude that the $P$-wave scattering in the spin quartet state has a key of the unfavorable 3NF contribution to $T_{21}$. 

The 3NF contribution to $T_{22}$ will be analyzed in a similar way. 
As seen in Fig.\ \ref{fig6}(b), $T_{22}^{[1]}$ gives small contributions to $T_{22}$.
Other two terms $T_{22}^{[2]}$ and $T_{22}^{[3]}$ are accompanied by the scalar amplitude $M^{(0)}_0(\frac32 \frac32)$ as Eqs.\ (\ref{eq:T2k_2}) and (\ref{eq:T2k_3}), which receives effectively no 3NF contribution as discussed above. 
The related tensor amplitude $M^{(2)}_{2}(\frac12 \frac32)$ and $M^{(2)}_{2}(\frac32 \frac32)$ are also scarcely affected  by the 3NF. 
In fact, $M^{(2)}_{2}(\frac32 \frac32)$ has been shown to be unaffected by the 3NF in neutron-deuteron scattering at $E_n=3$ MeV \cite{Is02}.  
However, $N_R$ in the denominator is affected by the 3NF, and most of 3NF contributions to $T_{22}$ are produced by this effect. 
Since $N_R$ is proportional to the cross section, the successful improvements by the 3NF in $d\sigma/d\Omega$ and $T_{22}$ are achieved due to the same origin, the central component of the 3NF.  

We will conclude that at $E_p=3.0$ MeV, the central part of the BR-3NF produces the successful contribution on the differential cross section as well as on the $T_{22}$ analyzing power, while the tensor part of the interaction gives the undesirable contribution to $T_{21}$.
The 3NF tensor effect on $T_{21}$ is not observed  below 1 MeV as seen in Table \ref{table1}, and Fig.\ \ref{fig6} suggests it may become appreciable above 2 MeV.
Thus precise measurements of the observables for the $pd$ scattering are highly desirable in such an energy region for the study of the 3NF tensor effect.
Also it will be interesting to examine the energy dependence of the effect up to higher energies, where several problems of the tensor analyzing powers have been reported \cite{Ki01b,Se02}.

\bigskip

This research was supported by the Japan Society for the Promotion of Science, 
under Grant-in-Aid for Scientific Research No. 13640300.
The numerical calculations was supported, in part, by the Computational Science 
Research Center, Hosei University, under Project No.\ lab0003.

\end{document}